# Non causal deep learning based dereverberation

*Jorge Wuth[1], Richard M. Stern[2], Nestor Becerra Yoma[1]*

[1]Speech Processing and Transmission Laboratory, Electrical Engineering Department,
University of Chile, Santiago, Chile
[2]Department of Electrical and Computer Engineering and Language Technologies Institute,
Carnegie Mellon University, Pittsburgh, PA 15206, USA

## ABSTRACT

In this paper we demonstrate the effectiveness of non-causal context for mitigating the effects of reverberation in deep-learning-based automatic speech recognition (ASR) systems. First, the value of non-causal context using a non-causal FIR filter is shown by comparing the contributions of previous vs. future information. Second, MLP- and LSTM-based dereverberation networks were trained to confirm the effects of causal and non-causal context when used in ASR systems trained with clean speech. The non-causal deep-learning-based dereverberation provides a 45% relative reduction in word error rate (WER) compared to the popular weighted prediction error (WPE) method in experiments with clean training in the REVERB challenge. Finally, an expanded multicondition training procedure used in combination with a semi-enhanced test utterance generation based on combinations of reverberated and dereverberated signals is proposed to reduce any artifacts or distortion that may be introduced by the non-causal dereverberation methods. The combination of both approaches provided average relative reductions in WER equal to 10.9% and 6.0% when compared to the baseline system obtained with the most recent REVERB challenge recipe without and with WPE, respectively.

## 1. INTRODUCTION

Many speech applications require that the user not be tethered to a close-talking microphone, such automatic meeting transcription, voice dialogue systems for devices in smart homes, and interaction with humanoid robots. In all of these cases the interactions are more intuitive, comfortable and effective if the user is able to interact with microphones on the device or at some third location, independent of the user. In many of these scenarios, the talker could be located several meters away from the microphone, and the received signal could be corrupted by interfering sounds, such as background noise and interfering speakers. In addition, speech in rooms is corrupted by the effects of reverberation caused by reflections of the speech from the surfaces of the room and the objects that are in it. The effects of reverberation are a major problem in distant-talking automatic speech recognition (ASR).

Reverberation (as well as background noise) decreases speech intelligibility and speech quality. This especially affects the performance of ASR systems, which are not as robust to reverberation as the human auditory system (Lippmann, 1997). These performance degradations depend on the nature of the environment and make such systems less effective (Furui, 2010); therefore, far-field speech recognition remains a challenge. One frequently-used measure of reverberation is the reverberation time (RT), which is defined as the time required for sound pressure level to decay by 60 dB. Offices and home environments typically have an RT from around 0.5 to 1.0 seconds, and longer RTs lead to greater reverberation distortion and greater degradation in ASR accuracy. Although RT is the most commonly-used general descriptor of reverberation, there are many other aspects of rooms that impact on speech intelligibility and ASR accuracy (Cowan, 2007).

As an example, Fig. 1 depicts the spectrograms of a clean utterance and the same utterance recorded in a reverberant room with an RT60 of approximately 0.9s. By comparing the two spectrograms, it can be noted that reverberation can be characterized by a smearing the spectrogram of clean speech toward the right (increasing running time) in the spectrogram depiction. Because of this characterization, the effects of reverberation in techniques such as non-negative matrix factorization (NMF) are frequently characterized by the convolution of the spectrogram with a causal linear filter. In this paper we consider the potential value in incorporating non-causal filtering into the compensation process.

There are several reasons why reverberation is an especially difficult problem for automatic speech processing systems. First, the impulse response of typical rooms is much greater than the durations typically used for analysis frames in speech recognition or speaker identification systems, which means that the spectral colorations introduced cannot be ameliorated by cepstral mean normalization or other frame-based noise-removal algorithms and must be treated over a longer time scale (*e.g.* (Gelbart & Morgan, 2002)). Many approaches based on adaptive noise cancellation that are quite successful in dealing with the effects of additive noise fail catastrophically in reverberant environments because the necessary assumption of statistical independence between the desired signal and the interfering source(s) is violated in reverberation, where the "distortion" consists of delayed and attenuated replications of the desired signal. Systems that are based on selective reconstruction based on time of arrival or relative intensity (*e.g*. (Roman, Srinivasan & Wang, 2006; Park & Stern, 2009; Kim et al., 2009)) fail because the reflections combine trigonometrically with the direct field and with each other to produce new amplitudes and phases that prevent the system from unambiguously inferring the direction of arrival of a sound source.

In recent years there has been increased interest in and attention to the development of systems that provide better robustness in reverberant environments, in part motivated by the REVERB challenge (Kinoshita et al., 2016) and the IARPA ASpIRE Challenge (Harper, 2015), both of which included challenging data in a heterogeneous set of reverberant and noisy environments. Results of the challenges were reviewed in the REVERB Workshop held in conjunction with IEEE ICASSP 2014 and a special session in IEEE ASRU 2015, respectively.

While it is not feasible to include a complete review and discussion of all approaches to reverberation remediation in this space, an excellent detailed analysis of the REVERB challenge may be found in (Kinoshita et al., 2016). It is noted in (Kinoshita et al., 2016) that systems that were successful in the REVERB challenge included a combination of enhancement at the waveform level and later, advanced acoustical modeling, typically using deep neural networks (DNNs), acoustic model adaptation, and occasionally system combination. Recent feature sets developed at least in part for reverberant environments include Damped Oscillator Coefficients (DOC) (Mitra, Franco & Graciarena, 2013), Normalized Modulation Coefficients (NMC) (Mitra et al., 2017), Modulation of Medium Duration Speech Amplitudes (MMeDuSA) features (Mitra et al., 2014), and Power Normalized Cepstral Coefficients (PNCC) (Kim & Stern, 2016). Some of the compensation algorithms that have been developed to cope with reverberation include various missing-feature approaches (*e.g*. (Cooke et al., 2001; Palomäki, Brown & Barker, 2004; Raj, Seltzer & Stern, 2004)), Suppression of Steady State and the Falling Edge (SSF) algorithm (Kim & Stern, 2010), non-negative matrix factorization (*e.g*. (Kumar et al., 2011)), and weighted prediction error (WPE) (Yoshioka et al., 2011; Yoshika, Chen & Gales, 2014). WPE performs dereverberation by processing the reverberated signal in the Short-Time Fourier Transform (STFT) domain by optimizing a causal FIR filter. As far as we know, a non-causal WPE-like filter has not been proposed in the literature yet.

### 1.1. Dereverberation with deep learning

Feedforward multilayer perceptrons (MLPs) have been employed for dereverberation by mapping spectral representations of reverberant speech frames to clean representations. Recognition accuracy, speech quality and intelligibility have usually been employed as metrics (Han et al., 2015; Han, Wang & Wang, 2014; Guzewich & Zahorian, 2017). Estimation of the spectral representation of the clean signal can use information from an individual reverberated frame or from a context window comprising more than one reverberated frame. For instance, in (Han et al., 2015) a context of 11 frames was employed, *i.e*. the center frames and five neighboring frames to each side, to estimate the target clean frame. Long short-term memory (LSTM) artificial neural network configurations have also been used extensively in dereverberation, leading to significant improvements in terms of word error rate (WER) (Mimura, Sakai, & Kawahara, 2015; Zhang et al., 2014; Weninger et al., 2014) and in signal-quality metrics such as STOI, PESQ or SNR gain (Mack et al., 2018; Zhao et al., 2018). To the best of our knowledge, the effects of non-causal context windows in the dereverberation process has not been addressed in MLP or LSTM frameworks.

### 1.2 Multi-condition training

Multi-condition training (MCT) (Seltzer, Yu & Wang, 2013) has been one of the most successful approaches to robust ASR and has also been employed to address the reverberation problem (Malek & Zdansky, 2019). MCT has been applied to several speech tasks, including speech recognition (Alam et al., 2015) and speaker verification (Peer, Rafaely & Zigel, 2008; Gammal & Goubran, 2005; Zhao, Wang & Wang, 2014; Couvreur & Couvreur, 2004; Avila et al., 2014). In MCT, speech distorted by various noise types, SNRs and reverberation levels is included in the dataset for acoustic model training. This method is usually performed in combination with data augmentation, where the training dataset is extended by artificially generating data. For example, clean speech may be convolved with a set of real or artificially-generated room impulse responses and added to noise that had been selected from various acoustical environments (*e.g.* (Ferras et al., 2016; Kim et al., 2017; Ko et al., 2017; Prisyach, Mendelev & Ubskiy, 2016)). Usually, MCT reduces the effect of distortion removal or enhancement techniques. We believe that adopting MCT strategies to take advantage of those distortion cancelling and enhancement methods has not received enough attention in the literature.

### 1.3 Goals of this study

In this paper we consider and evaluate various approaches that exploit non-causal context for deep-learning-based dereverberation. First, we consider the potential value of using a non-causal FIR filter to make use of non-causal context in the incoming signals specifically comparing the contributions of previous (non-causal) versus future (causal) information in estimating clean speech from reverberant speech. Second, MLP-based and LSTM-based dereverberation systems are trained to assess the effect of causal versus non-causal context using ASR experiments with clean training data. As we will show, the use of non-causal deep-learning-based dereverberation provides a 45% reduction in WER compared to WPE in experiments with clean training in the REVERB challenge. It is worth highlighting that ASR based on clean training was adopted for the initial evaluation because the objective of the dereverberation methods is to reduce the mean squared error (MSE) between the reverberated signal and a clean reference. Finally, we discuss the potential impact of various approaches to ASR acoustic-model training that incorporate MCT. The most successful approach makes use of enhanced training data that includes both clean and reverberated speech to reduce potential artifacts or distortions that may be introduced by the non-causal dereverberation method. The combination of these approaches provides average reductions in WER equal to 10.9% and 6.0% when compared to the baseline system provided by the most recent REVERB challenge recipe without and with WPE, respectively.

# 2. DEREVERBERATION AND NON-CAUSALITY

Our dereverberation algorithms operate on the STFT $X[n,k]$ of the reverberated signal $x(t)$. We consider the trajectory $\boldsymbol{X}_k = [X_k(1), \dots X_k(n) \dots, X_k(N)]$ of the $k$-th frequency bin of $X[n,k]$ along $N$ frames, where $n$ corresponds to the time-frame index. The effect can also be seen in the autocorrelation of this trajectory for the clean and reverberated utterances. Figure 2 compares the average normalized autocorrelation of all frequency bins and time trajectories estimated using the clean testing utterances from the AURORA-4 database and the corresponding reverb-test-set. These two datasets are described in Sec. 2.2. It can be seen that the reverberated speech exhibits greater autocorrelations in the tails than clean speech. This implies that a dereverberation method must decrease the autocorrelation of the trajectories of the reverberated utterances in each frequency bin. Nevertheless, we note that while a dereverberation process must produce less correlated bin trajectories, the converse is not true, as more decorrelated trajectories do not necessarily corresponds to a proper cancelation or attenuation of the reverberation effect. Decorrelation of the trajectory is a necessary but not sufficient condition for dereverberation.

### 2.1. Non-causal MSE clean signal estimation

To study the contribution of non-causality, the dereverberation process is modeled as a linear filter that operates on the reverberated signal and outputs an estimate of the clean signal. Specifically, we seek to obtain a filter $\boldsymbol{g}_k = [g_k(1), \dots, g_k(q+1), \dots, g_k(p+q+1)]$ that when applied over the trajectory $\boldsymbol{X}_k$ of the reverberated signal $x(t)$ returns an estimate of the corresponding clean signal trajectory, $\boldsymbol{Y}_k = [Y_k(1), \dots Y_k(n) \dots, Y_k(N)]$. This can be formulated as follows:

$$\hat{Y}_k(n) = g_k(1)X_k(n+q) + \cdots + g_k(q+1)X_k(n) + \cdots + g_k(p+q+1)X_k(n-p) \qquad (1)$$

where $\widehat{\boldsymbol{Y}}_{\boldsymbol{k}} = [\hat{Y}_k(1), \dots \hat{Y}_k(n) \dots, \hat{Y}_k(N)]$ corresponds to the estimated $k$-th frequency bin trajectory of the clean signal; $1 \leq n \leq N-q$, $N$ is the length of signal $x(t)$ in frames; and $X_k(n) = 0$ when $n < 1$. Note that $p$ and $q$ represent the number of previous $p$ samples, (*i.e.* causal context), and future $q$ samples, (*i.e.* non-causal context), respectively, that are considered in estimating the current frame of dereverberated speech. Filter $\boldsymbol{g}_k$ is obtained by solving the following optimization problem:

$$\boldsymbol{g}_k = \underset{\boldsymbol{g}}{\operatorname{argmin}} \sum_{n=1}^{N_c} \left|\hat{Y}_k(n) - Y_k(n)\right|^2 \qquad (2)$$

where $\boldsymbol{g}_k = \boldsymbol{g}_{r_k} + j\boldsymbol{g}_{j_k}$, and $\boldsymbol{g}_{r_k}$ and $\boldsymbol{g}_{j_k}$ correspond to the real and imaginary components of $\boldsymbol{g}_k$**,** respectively;

$N_c$ is the length of the clean signal $y(t)$ in frames; and $E = \sum_{n=1}^{N_c} \left| \hat{Y}_k(n) - Y_k(n) \right|^2$ corresponds to the prediction error that we seek to minimize. As explained in the Appendix, Eq. (2) yields the following solutions with index $k$ removed to simplify the notation:

$$\boldsymbol{g_r} = \left( \left(\boldsymbol{M_{rr}} + \boldsymbol{M_{jj}}\right)^{-1} \left(\boldsymbol{M_{jr}} - \boldsymbol{M_{rj}}\right) - \left(\boldsymbol{M_{rj}} - \boldsymbol{M_{jr}}\right)^{-1} \left(\boldsymbol{M_{rr}} + \boldsymbol{M_{jj}}\right) \right)^{-1} \left( \left(\boldsymbol{M_{rr}} + \boldsymbol{M_{jj}}\right)^{-1} \left(\boldsymbol{R_{X_jY_r}} - \boldsymbol{R_{X_rY_j}}\right) - \left(\boldsymbol{M_{rj}} - \boldsymbol{M_{jr}}\right)^{-1} \left(\boldsymbol{R_{X_rY_r}} - \boldsymbol{R_{X_jY_j}}\right) \right) \quad (3)$$

$$\boldsymbol{g_j} = \left( \left(\boldsymbol{M_{rr}} + \boldsymbol{M_{jj}}\right)^{-1} \left(\boldsymbol{M_{rj}} - \boldsymbol{M_{jr}}\right) - \left(\boldsymbol{M_{jr}} - \boldsymbol{M_{rj}}\right)^{-1} \left(\boldsymbol{M_{rr}} + \boldsymbol{M_{jj}}\right) \right)^{-1} \left( \left(\boldsymbol{M_{jr}} - \boldsymbol{M_{rj}}\right)^{-1} \left(\boldsymbol{R_{X_jY_r}} - \boldsymbol{R_{X_rY_j}}\right) - \left(\boldsymbol{M_{rr}} + \boldsymbol{M_{jj}}\right)^{-1} \left(\boldsymbol{R_{X_rY_r}} - \boldsymbol{R_{X_jY_j}}\right) \right) \quad (4)$$

Refer to the Appendix for the definition of $\boldsymbol{M_{rr}}$, $\boldsymbol{M_{jj}}$, $\boldsymbol{M_{rj}}$, $\boldsymbol{M_{jr}}$, $\boldsymbol{R_{X_rY_r}}$, $\boldsymbol{R_{X_jY_j}}$, $\boldsymbol{R_{X_rY_j}}$, $\boldsymbol{R_{X_jY_r}}$. Different causal and non-causal context lengths were employed to evaluate the benefits of considering a non-causal filter, and to study the effect of different context lengths.

### 2.2. Reverberated data generation with AURORA-4 database

The reverberated data used to evaluate the performance of the non-causal filter and the deep-learning-based dereverberation approaches studied here were derived from the AURORA-4 database (Hirsch, 2002), which in turn was generated from the 5000-word closed-loop vocabulary task based on the DARPA Wall Street Journal (WSJ0) Corpus. The AURORA-4 corpus is divided into 3 sets, namely, training, development, and evaluation sets. The training set contains 7,138 utterances from 83 speakers, totaling 14 hours of speech data. This database was used to allow comparison with results published by the speech community. The reverberated data were obtained by convolving the clean utterances of the AURORA-4 corpus with artificially generated room impulse responses (RIRs). The RIRs were simulated using the Room Impulse Response Generator (Habets, 2010). The reverberation time (RT) values of the generated RIRs varied between 0.4 and 1.99 seconds. The dimensions of the virtual room used to generate each individual RIR were drawn from uniform distributions over the range of plus or minus 20 percent of the nominal values of 7.95 m length, 5.68 m width and 4.5 m height, which are the approximate dimensions of a typical non-rectangular meeting room as the ones present in our department. The speaker-to-microphone distance was drawn from a uniform distribution between 0.144 and 2.816 m. The speaker and microphone were placed in random locations at the room that were selected individually for a particular trial, with the constraints that both speaker and microphone were at least 1 m from any wall and between 1 m and 2 m from the floor. This randomization of the simulation parameters was implemented to reduce potential effects of artifacts caused by standing-wave phenomena in the rectangular shoebox-shaped room that RIR and other similar simulations based on the image method (Allen & Berkley, 1979). By using this procedure, a set of 8,000 RIRs was obtained.

Three sets of reverberated data were derived from the AURORA-4 corpus, *i.e.*, a training, development, and evaluation sets. These sets were generated by convolving each utterance of the corresponding clean set with a simulated room impulse response (RIR) selected randomly from the list of generated RIRs. The three sets were generated in such a way that there were no two utterances were convolved with the same RIR. These set will be referred to as reverb-train-set, reverb-dev-set, and reverb-test-set. The original clean sets will be referred to as clean-train-set, clean-dev-set, and clean-test-set.

### 2.3. Speech recognition experiments with AURORA-4

We evaluated the impact of different context lengths in the non-causal filter, as well as the effect of other deep-learning-based dereverberation approaches in terms of WER by means of speech recognition experiments with a DNN-HMM system built with clean data using the Kaldi Speech Recognition Toolkit (Povey et al., 2011). Specifically, we constructed a DNN-HMM ASR system using the tri2b Kaldi recipe for the AURORA-4 database. The system was trained with the original clean data from the database. First a GMM-HMM was built by training a monophone system; then, the alignments from that system were employed to generate an initial triphone system; finally, the triphone alignments were employed to train the final triphone system. This recipe also included mel-frequency cepstral coefficients (MFCCs), linear discriminant analysis (LDA), and maximum likelihood linear transforms (MLLTs). Once the GMM-HMM system was trained, the GMM was replaced by a DNN. The DNN was composed of seven hidden layers and 2048 units per layer each, and the input considered a context window of 11 frames. The number of units of the output DNN layer was equal to the number of Gaussians in the corresponding GMM-HMM system. The reference for the DNN training was the alignment obtained with the training data and the GMM-HMM. The feature vector consisted of 40 Mel filter bank (MelFB) features, and delta and delta-delta dynamic features, using an 11-frame context window. The DNN was trained initially using the Cross-Entropy criterion. Then, the final system was obtained by re-training the DNN using sMBR discriminative training (Veselý et al., 2013). For decoding, the standard 5K lexicon and trigram language model from WSJ were used (Gauvain, Lamel & Adda-Decker, 1995). As a result, the language model was tuned to the task, *i.e*. it is task dependent.

## 3. NON-CAUSAL DEREVERBERATION WITH DEEP LEARNING

We tried to modify the original WPE formulation by replacing the ordinary causal filter with a non-causal one.

However, by following the procedure described in (Nakatani et al., 2010), the likelihood function obtained with our non-causal filter could not be rewritten as shown in Appendix C of their paper. As a result, we could not follow the same procedure to estimate an equivalent minimum norm solution as reported in (Nakatani et al., 2010) for the optimization procedure. For these reasons, we adopted deep learning-based approaches to evaluate the pertinence and usefulness of non-causal context information for dereverberation.

Each utterance was divided into 25-ms time frames with 10-ms frame shift, and a 512-point fast Fourier transform (FFT) was applied to compute spectral magnitudes in each time frame, resulting in a total of 257 non-negative frequency bins. 40 Mel filterbank coefficients were extracted from the STFT spectral magnitudes in each frame, and the logarithm was applied to each filter energy. By doing so, each frame can be represented as a vector $\boldsymbol{LFE}(n)$:

$$\boldsymbol{LFE}(n) = [LFE(n,1), LFE(n,2), \dots, LFE(n,40)]^T \tag{5}$$

where $LFE(n,k)$ corresponds to the log filter energy of filter $k$ in frame $n$. Finally, we applied mean and variance normalization (MVN) for each one of the 40 log filter energy trajectories along the utterance on a per-utterance basis. In order to evaluate the benefits of temporal context, we incorporated the spectral features of neighboring frames into the feature vector of the reverberated signal. Therefore, the reverberated feature vector is:

$$\widetilde{\boldsymbol{LFB}_{\boldsymbol{X}}}(n) = [LFB_X\,(n-p)^T, \dots, LFB_X(n)^T, \dots, LFB_X(n+q)^T]^T \tag{6}$$

where $p$ and $q$ denote the number of causal and non-causal context frames, respectively. Consequently, the dimensionality of the reverberated feature vector is $(p+q+1)$ x 40. The clean feature vector is the current clean frame $n$, denoted by a 40-dimensional feature vector $\boldsymbol{LFE_Y}(n) = [LFE_Y(n,1), LFE_Y(n,2), \dots, LFE_Y(n,40)]^T$ as defined in Eq. (5). The estimated clean feature vector corresponds to $\boldsymbol{LFE_{\hat{Y}}}(n) = [LFE_{\hat{Y}}(n,1), LFE_{\hat{Y}}(n,2), \dots, LFE_{\hat{Y}}(n,40)]^T$. Two deep-learning-based approaches were evaluated to remove the reverberation effect, MLP and LSTM, and the dimensional reduction from 257 FFT bins to 40 Mel filters allowed to reduce the computational load of the deep learning training procedures. The deep-learning-based dereverberation approaches were implemented and tested on a GPU using the Tensorflow open source software library (Abadi et al., 2015).

### 3.1. MLP-based dereverberation

An MLP was trained to learn the mapping from the reverberant signal features to the clean ones (see Fig. 3). The training samples corresponded to the reverberated and clean utterances that were parametrized as described above. For each frame $n$ of a given reverberated utterance, the input was the reverberated feature vector $\widetilde{\boldsymbol{LFB}_{\boldsymbol{X}}}(n)$, and the number of MLP input units was the same as the dimensionality of this feature vector. The reference was the feature vector $\boldsymbol{LFE_Y}(n)$ corresponding to the corresponding clean frame $n$. The objective function for optimization was based on the mean squared error (MSE) between the output of the MLP and the corresponding clean feature vector $\boldsymbol{LFE_Y}(n)$. The activation function in the hidden layers was the sigmoid function and no activation function was employed in the output layer. The weights of the MLP were initialized with pretraining. We used backpropagation with a mini-batch for stochastic gradient descent to train the MLP model, and the cost function in each mini-batch is computed as the summation over multiple training samples. The optimization technique used gradient decent along with adaptive learning rates. The MLP hyperparameters were determined empirically based on WER with the clean DNN-HMM system described above while keeping the number of causal and non-causal equal to 40. These parameters corresponded to the number of layers of the MLP, hidden units per layer, learning rate and batch size. Then, additional speech recognition experiments were performed to determine the optimal number of causal and non-causal frames, $p$ and $q$, respectively.

### 3.2. LSTM-based dereverberation

An LSTM neural network was also trained to learn the mapping from the reverberant signal features to the features representing the clean speech. Given a reverberated utterance, all its feature vectors $\widetilde{\boldsymbol{LFB}}_{\boldsymbol{X}}(n)$, with n$= 1, \dots, N_c$, where $N_c$ is the number of frames of the corresponding clean signal, were processed by the LSTM neural network at a time. The duration of the reverberated signal is usually slightly greater than that of the clean signal and the minimization of the cost function requires a one-to-one correspondence between reverb and clean feature vectors. Hence, the dimension of the input matrix for each training utterance was $K$ **x** $N_C$, where $K$ is the dimension of each feature vector $\widetilde{\boldsymbol{LFB}}_{\boldsymbol{X}}(n)$, *i.e.* $K = (p + q + 1)$ x $40$. The output corresponded to all the feature vectors $\boldsymbol{LFE_Y}(n)$, where $n = 1, \dots, N_c$, of the clean signal, and the dimension of the output matrix was $40$ x $N_c$. The objective function for optimization was also based on the MSE between the output of the LSTM and the corresponding clean utterance feature vectors. The weights of the LSTMs were randomly initialized without pretraining. We used backpropagation with a mini-batch of one for stochastic gradient descent. The optimization technique uses gradient descent along with adaptive learning rates. Three different LSTM architectures where evaluated: unidirectional forward, unidirectional backward, and bidirectional. These networks are shown in Figs. 4 and 5. First, the number of causal and non-causal frames, $p$ and $q$, were kept fixed and equal to 0 while the best architecture and hyperparameters were determined based on WER with the clean DNN-HMM system. Experiments were performed to determine the best architecture and, with the chosen architecture, we optimized other hyperparameters including the number of output units per LSTM, and the learning rate. Finally, additional speech recognition experiments were performed to determine the optimal number of causal and non-causal frames, $p$ and

$q$, respectively.

## 4. DEREVERBERATION AND MULTICONDITION TRAINING

The non-causal deep-learning-based dereverberation approaches were applied to the REVERB challenge corpus (Kinoshita et al., 2016) for validation purposes. The REVERB challenge provides a training set, a development (Dev) test set, and an evaluation (Eval) test. The Dev and Eval test sets are composed of real recordings (RealData) and simulated data (SimData). SimData refers to reverberant utterances generated based on the WSJCAM0 corpus (Robinson et al., 1995) by convolving clean signals with measured room impulse responses (RIRs) and subsequently adding measured stationary ambient noise signals with a signal-to-noise ratio (SNR) of 20 dB. SimData simulated six different reverberation conditions: three rooms with different volumes (small, medium, and large) and two distances between a speaker and a microphone array (near and far, which were 50 and 200 cm, respectively). The reverberation times (T60) of SimData-room1, -room2, and -room3 are about 0.3s, 0.6s, and 0.7s, respectively. RealData consists of utterances spoken by human speakers in a noisy and reverberant room. The room's T60 was about 0.7 s (Lincoln et al., 2005). The recordings contain some stationary ambient noise. RealData contains two reverberant conditions: one room and two distances between the speaker and the microphone array (near and far, which were ∼100 and ∼250 cm, respectively). The six conditions in SimData and the two conditions in RealData make up eight testing conditions. The training dataset consists of (i) a clean training set taken from the original WSJCAM0 training set and (ii) an MCT set generated from the clean WSJCAM0 training data by convolving the clean utterances with 24 measured room impulse responses and adding recorded background noise at an SNR of 20 dB. In this paper, the eight 1-ch testing conditions were considered. In addition, different MCT procedures were proposed and evaluated here.

### 4.1. Non-causal MLP- and LSTM-based dereverberated training

By using the procedures described in Secs. 3.1 and 3.2, a non-causal MLP with $p$=10 (causal context) and $q$=10 (non-causal context) and a non-causal LSTM with $p$=10 and $q$=5 were trained using the clean training and the 8-ch MCT sets of the REVERB challenge. The training samples correspond to each one of 62,888 reverberated signals of the REVERB challenge MCT training set (7,861 clean utterances convolved with eight channels) and their corresponding clean features vectors from the 7,861 clean utterances that comprises the clean training set. The hyperparameters of the network, as well as the causal and non-causal context length, corresponded to those obtained empirically in Secs. 3.1 and 3.2 with the AURORA-4 database.

### 4.2. TDNN-based ASR

TDNN systems were built based on the most recent Kaldi recipe for the REVERB challenge[1]. To build these TDNN systems, first a GMM-HMM system was trained. This GMM-HMM system was used to force-align the training data or an augmented version of it to produce alignments to train the TDNN which employs the lattice-free maximum mutual information (LF-MMI) objective function (Povey et al., 2016). The input to the TDNN corresponds to the log magnitude of the 40 Mel filter bank features with MVN (see Sec. 3) at each time step and appended to a 100-dimensional iVector (Dehak et al., 2010). Consequently, the TDNN input is a 140-dimensional feature vector. For decoding, the standard 5K lexicon and trigram language model from WSJ were used (Gauvain, Lamel & Adda-Decker, 1995). The REVERB challenge recipe was run on a machine with an Intel® Core™ i7-7700 CPU, 32GB of RAM, and 1 NVIDIA GTX1080 GPU, which is different from the hardware mentioned by the recipe that assumes from 2 to 4 GPUs as a default. Because of these restrictions, the system training was modified somewhat, following guidance by Povey and Trmal (personal communication, 2019). Also, the original cepstral input for the TDNN was replaced with the log magnitude of 40 Mel filter bank features with MVN as described in Sec. 3 because this input vector provided better results. The Kaldi version used in this paper corresponds to 5.5.576 from December 12th, 2019.

The different training conditions evaluated here are described as follows.

#### *4.2.1. Clean training*

For clean training an ASR system was trained by using the 7,861 utterances from the clean training set provided with the REVERB challenge. Both the GMM-HMM and TDNN were trained with the clean data.

#### *4.2.2. Multicondition training (MCT)*

For the experiments using MCT, a GMM-HMM was trained by using the 62,888 utterances from the 8-ch MCT set provided with the REVERB challenge. The TDNN was trained by using an augmented version of the 8-ch MCT training set. The training data were augmented by generating two additional copies corresponding to speed perturbations of 0.9 and 1.1 according to the recipe, increasing the amount of data for training the TDNN to 188,664 utterances. Additionally, the amplitude of each recording in the training data was scaled by a random variable drawn from a uniform distribution over $\left[\frac{1}{8}, 2\right]$.

#### *4.2.3. Expanded multicondition training: MLP and LSTM*

We also propose in this paper an expanded form of MCT in which the TDNN system was trained with three training datasets: the clean training set; the 8-ch MCT set as well as a dereverberated version of the 8-ch MCT training set. The motivation for expanding the data for this MCT was to take advantage of the MSE reduction with respect to the clean signal led by the non-causal dereverberation deep learning schemes and to minimize any artifact or distortion introduced by these enhancement techniques.

[1] https://github.com/kaldi-asr/kaldi/tree/master/egs/reverb/s5

### 4.3. Reverberated, dereverberated and semi-enhanced test data

The eight 1-ch testing conditions from the REVERB challenge were dereverberated by using the non-causal MLP and LSTM schemes trained as described in Sec. 4.1. Additionally, in this paper we consider a semi-enhanced version of the testing data by linearly combining the log Mel filter energy trajectories from the reverberated utterances and the corresponding dereverberated utterances. The motivation is the same as the expanded MCT procedure discussed above: to make proper use of the decrease in MSE with respect to the clean signal caused by the non-causal dereverberation deep learning schemes and to minimize the artifact introduced by these enhancement techniques.

In our experiments we compare results obtained using four different mixtures of features which differ in where the features are extracted and how features with and without WPE are combined. We let $\boldsymbol{LFE_{SE}}(n) = [LFE_{SE}(n,1), LFE_{SE}(n,2), \dots, LFE_{SE}(n,40)]^T$ denote the feature representation of frame $n$ of a semi-enhanced testing signal, the mixing parameter $\lambda$ a constant between zero and one. $WPE(.)$ indicates that WPE was applied to a given utterance, and $\boldsymbol{X}$ and $\widehat{\boldsymbol{Y}}(\boldsymbol{X})$ are the reverberated and the corresponding non-causal deep-learning-based dereverberated signals. The four configurations used to generate the semi-enhanced test utterances are described as follows.

**Configuration 1-** In this configuration, WPE is applied to the reverberated utterance and the dereverberated signal is obtained from the reverberated signal using the non-causal MLP or LSTM schemes:

$$\boldsymbol{LFE_{SE}}(n) = (1-\lambda)\cdot WPE[\boldsymbol{LFE_X}(n)] + \lambda\cdot \boldsymbol{LFE_{\widehat{Y}(X)}}(n) \tag{7}$$

**Configuration 2-** In this configuration, WPE is applied to the reverberated utterance and the dereverberated signal is obtained using a non-causal MLP or LSTM from the reverberated signal also after being processed using WPE:

$$\boldsymbol{LFE_{SE}}(n) = (1-\lambda)\cdot WPE[\boldsymbol{LFE_X}(n)] + \lambda\cdot \boldsymbol{LFE_{\widehat{Y}[WPE(X)]}}(n) \tag{8}$$

**Configuration 3-** In this configuration, the reverberated utterance is combined with a dereverberated signal obtained using a non-causal MLP or LSTM from the reverberated signal after being processed using WPE:

$$\boldsymbol{LFE_{SE}}(n) = (1-\lambda)\cdot \boldsymbol{LFE_X}(n) + \lambda\cdot \boldsymbol{LFE_{\widehat{Y}[WPE(X)]}}(n) \tag{9}$$

**Configuration 4-** In this configuration, the reverberated utterance is combined with the dereverberated signal obtained using a non-causal MLP or LSTM from the reverberated signal:

$$\boldsymbol{LFE_{SE}}(n) = (1-\lambda)\cdot \boldsymbol{LFE_{X}}(n) + \lambda \cdot \boldsymbol{LFE_{\hat{Y}(X)}}(n) \qquad (10)$$

## 5. RESULTS AND DISCUSSION

In this section we describe and discuss the results of the experiments that were performed to evaluate the impact on recognition accuracy of the various techniques introduced in the sections above. We begin with a series of initial results and continue with discussions of the impact of expanded MCT, semi-enhanced test utterances, and a combination of these two techniques.

### 5.1. Initial evaluations

We first discuss results using the filter described in Sec. 2.1 implemented with different context lengths. The evaluation of the causal and non-causal context length was performed by ASR experiments using a DNN-HMM system trained with clean data (see Sec. 2.3) from the AURORA-4 database. ASR experiments using clean training were adopted for the initial evaluations because the dereverberation methods were designed to reduce the MSE between the reverberated signal and the clean reference. The WERs obtained with clean and reverberated testing utterances are, respectively, 2.2% and 76.6% (Table 1). The results for acoustic models obtained by training with clean speech for different causal and non-causal context lengths are shown in Figs. 6-9.

The MLP (Fig. 3) and LSTM (Figs. 4-5) hyperparameters were tuned as described in Sec. 3.1. The optimal MLP configuration corresponded to three hidden layers with 1000 nodes for each hidden layer, trained using a learning rate and batch size equal to 0.1 and 200, respectively. In the case of the LSTM network, the optimal configuration is defined by the bidirectional architecture shown in Fig. 5, including 100 output units per LSTM and trained with a learning rate equal to 0.1.

Figure 6 shows the dependence of the minimal MSE on the number of frames of causal and non-causal context. It can be seen that the global best performance is obtained when the clean signal is estimated as in Eq. (1) using a non-causal polynomial that makes use of future and previous reverberated samples. Accordingly, Fig. 7 shows that, for a given number of free parameters in Eq. (1), the lowest WERs are obtained when the causal/non-causal context ratio is around 50%, *i.e.* when the number of future and previous samples are approximately equal. Figure 8 shows the spectrograms of a clean, reverberated, and dereverberated utterance using Eq. (1) with this equation derived in detail in the Appendix. We note that the blurring effect of the reverberation is significantly removed. The autocorrelation function for FFT bin trajectories provided by a clean, reverberated, and dereverberated utterance for four different utterances is shown in Fig. 9. As expected, the MSE-based non-causal dereverberation fitting scheme according to Eqs. (1) and (2) leads to less correlated FFT bin trajectories.

ASR results obtained using a DNN-HMM system with the AURORA-4 database and simulated reverberation are presented in Figs. 10 and 11. Figure 10 shows the WER obtained with the non-causal dereverberating MLP (Fig. 3). We note that these optimal parameter values were not obtained using the fitting procedure defined by Eqs. (1) and (2). Instead, the enhancement MLP is trained and testing using different utterances and hence is forced to generalize. The WER curve in Fig. 10 has many local minima, although it clearly suggests that lowest WERs are obtained with a small number of combined future and previous samples rather than a greater number of previous samples only. Figure 11 depicts the corresponding WERs obtained with the non-causal dereverberating bidirectional LSTM (Fig. 5). We observed that the lowest WERs are obtained when the neighborhood window incorporates both future (non-causal context) and previous (causal context) samples to estimate the clean signal trajectory, and best results were obtained using causal and non-causal contexts with $p = 10$ and $q = 10$ for the MLP, and $p = 10$ and $q = 5$ for the LSTM. These results are consistent with the trends of Fig. 7.

Figure 12 shows the 40-log filter energy spectrograms of a clean utterance, along with the corresponding reverberated and dereverberated version of it. The log filter energy trajectories were mean and variance normalized. The reverberated signal was generated by convolving the clean utterance with a generated RIRs with RT of 0.8 seconds. The dereverberated signals were obtained with the MLP and LSTM networks using their optimal parameters. It can be seen that the dereverberating MLP and LSTM networks significantly reduce the blur towards the right in the spectrogram produced by reverberation.

Table 1 summarizes the WERs obtained using the initial system and the AURORA-4 database along with the optimal parameters for the MLP and LSTM networks, as described above. We note that with clean training, the application of dereverberation using the MLP and especially the LSTM networks leads to much better performance on the reverberated test data than was obtained with WPE, providing average relative reductions in WER as high as 61% and 49% when compared to the baseline system and WPE, respectively, using the AURORA-4 database. The fact that the LSTM-based dereverberation scheme provides lower WERs in general than the MLP-based enhancement may be suggest that the recurrent nature of the LSTM network is more suitable to modeling reverberation.

The remainder of the experimental results in this paper were obtained using data from the REVERB training and the TDNN ASR described in Sec. 4.2, which was our best approximation of the TDNN system provided by the REVERB challenge given the differences in our computing environment. Results for the REVERB challenge using clean training are presented in Table 2. The dereverberating MLP and LSTM configurations correspond to those that provided the lowest WERs with the AURORA-4 database as described above. The WERs obtained with clean utterances are similar and competitive to those published elsewhere. When compared to the baseline experiment with the reverberated utterances, the average relative improvement in WER provided by the MLP- and LSTM-based schemes was equal to 53.0%. When compared to WPE, the average relative improvement in WER given by the dereverberating MLP and LSTM was equal to 45.7%. These results basically corroborate the utility of incorporating non-causal context samples to remove the effects of reverberation.

Table 3 shows the corresponding results for the REVERB challenge and MCT. Surprisingly, when multicondition training applied according to the REVERB challenge recipe, the reverberating MLP and LSTM did not outperform the baseline system neither without or with WPE. These results were unexpected considering that both the enhancing non-causal MLP and LSTM dramatically reduces the MSE with the respect to the clean signal. This outcome may have been due to the fact that the baseline TDNN system provided by the REVERB challenge recipe models the reverberated speech very well. Also, the dereverberating deep-learning-based schemes studied here may have introduced a distortion that is not represented properly by the TDNN baseline system. Finally, MCT leaves little room for improvement considering the results with clean speech obtained with the ASR system trained with clean utterances as shown in Table 2. In any case, these disappointing results motivated us to consider alternate approaches that reformulated the training procedure and the generation of testing utterances that would take advantage of the enhancement effect obtained by the dereverberating non-causal MLP and LSTM. We describe this work in the sections that follow.

**5.2. Impact of expanded MCT.**

The expanded MCT approach consists of training the TDNN system with three training datasets: the clean training set; the 8-ch MCT set as well as a dereverberated version of the 8-ch MCT training set, as described in Sec. 4.2.3 above.

Results for the REVERB challenge and the expanded MCT are presented in Tables 4 and 5, representing dereverberation using the MLP and LSTM networks, respectively. As seen in Tables 4 and 5, the use of expanded MCT provides substantially better WER than the use of the original MCT as described in Tables 2 and 3. Nevertheless, even with expanded MCT, the recognition accuracy is not better than what is already provided by WPE.

**5.3. Impact of semi-enhanced test utterance generation combined with enhanced MCT**

As discussed in Sec. 4.3, this approach describes the use of a linear combination of features from the original reverberant utterance with features obtained after the dereverberation process. Four configurations are specified, which differ in how and where WPE is applied. In addition, the value of the optimal mixing parameter $\lambda$ depends on the configuration and subset of testing data considered. In our application of these approaches, Eqs. (7), (8),

(9) and (10) were tuned for each REVERB challenge Dev data subset, producing subset-dependent $\lambda\text{'}s$ for each semi-enhanced testing configuration. Additionally, an average optimal $\lambda$ was computed for each semi-enhanced testing. It is worth noting that while the ASR system is trained in multicondition fashion, we continued to use MLP and LSTM networks for dereverberation that were trained to learn the mapping from the features of the reverberant speech to clean speech as described in Sec. 3.

Tables 6, 7, 8, and 9 describe results obtained when semi-enhanced test utterance generation is combined with the expanded MCT procedure and non-causal LSTM-based dereverberation. Tables 6 and 7 describe results obtained using the MLP dereverberation network, while Tables 8 and 9 provide the corresponding results using the LSTM dereverberation network. Table 6 and 8 describe results in which the λ parameter is optimized for each REVERB challenge testing subset, while in Tables 7 and 9, the average optimal $\lambda$ was employed for all the REVERB challenge testing subsets. The average optimal values of $\lambda$ for Configurations 1,2,3 and 4 using the MLP network were 0.3, 0.363, 0.425 and 0.425, respectively, while the corresponding best values of λ using the LSTM network were 0.2625, 0.3375, 0.4 and 0.4, respectively.

As can be seen in Table 6, the use of an optimal $\lambda$ for each REVERB challenge testing subset, provided average relative reductions in WER equal to 1.2%, 1.7% and 1.5% for Configurations 1, 2 and 3 respectively, when compared to the baseline system with WPE. This approach did not provide any reduction in WER for Configuration 4. In Table 7, a single average optimal $\lambda$ was employed for all the REVERB challenge testing subsets. The use of the average optimal $\lambda s$, provided average relative reductions in WER equal to 2.4%, 2.8%, 0.6%, respectively, for Configuration 1, 2, 3 and 4, when compared to the baseline system with WPE. We note that the non-causal MLP was not re-trained with the log filter energies after being processed by WPE.

As can be seen in Table 8, where an optimal $\lambda$ was employed per each REVERB challenge testing subset, Configurations 1,2,3 and 4, corresponding to Eqs. (7), (8), (9) and (10), respectively, led to average relative reductions in WER equal to 4.2%, 4.9%, 2.8% and 1.7%, respectively, when compared to the baseline system with WPE. In Table 9, the average optimal $\lambda$ was employed for all the REVERB challenge testing subsets. The use of these average optimal $\lambda$´s in Configurations 1,2,3 and 4 provided an average relative reduction in WER equal to 3.6%, 6.0%, 3.9% and 2.1%, respectively, when compared to the baseline system with WPE. We note once again that the non-causal LSTM was not re-trained with the log filter energies after being processed by WPE. As a reference, an oracle improvement can be obtained if $\lambda$ is tuned for each REVERB challenge test data subset for Configuration 2, which led to the best results. By doing this, the relative reductions in WER is equal to 8.3% when compared to the baseline system with WPE. Additionally, if a constant $\lambda = 0.5$ is used, a relative reduction in WER equal to 5.8% is observed when compared to the baseline system with WPE.

Summarizing results in Tables 6-9, the combination of the reverberated and dereverberated signals, *i.e*. the proposed semi-enhanced test utterance generation, provides significant reductions in WER when applied in combination with expanded multicondition training as described above. In this condition, combining the reverberated and enhanced utterances takes advantage of the original distortion reduction resulting from the enhancing method by reducing the effect of any artifact introduced by the dereverberation scheme. It is also worth noting that the non-causal dereverberation MLP and LSTM still provides improvements in recognition accuracy even when applied after WPE without any further retraining. This result may be due to the fact that WPE employs a causal model and the non-causal dereverberation makes use of information that is not being employed.

## 6. CONCLUSIONS

We demonstrate in this paper the effectiveness and utility of non-causal context for deep-learning-based dereverberation. First, we evaluated the impact of non-causal context using a non-causal FIR filter, specifically comparing the contributions of previous vs. future information. Second, dereverberating MLP- and LSTM-based systems were trained to assess the effects of causal/non-causal context by making use of ASR experiment with clean utterances from AURORA-4 database. The non-causal deep-learning-based dereverberation can lead to a 45% relative reduction in WER compared to WPE in experiments with clean training in the REVERB challenge. Finally, we proposed an expanded multicondition training procedure and a semi-enhanced test utterance generation based on combinations of reverberated and dereverberated signals to reduce any artifacts or distortion that may be introduced by the non-causal dereverberation method. The combination of both approaches can lead to average relative reductions in WER equal to 10.9% and 6.0% when compared to the baseline system provided by the most recent REVERB challenge recipe without and with WPE, respectively. The evaluation of the proposed expanded multicondition training method and the semi-enhanced test utterance generation scheme with other enhancement or distortion removal methods in speech related recognition problems can be considered in future research.

## APPENDIX

Using Eq. (1), Eq. (2), can be rewritten as

| | |
|---|---|
| $$\boldsymbol{g}^* = \underset{\boldsymbol{g}}{\operatorname{argmin}} \sum_{n=1}^{N_c} \left| \sum_{l=1}^{p+q+1} g(l)X(n-l+q+1) - Y(n) \right|^2$$ | (A1) |

where the subscript $k$ corresponding to the bin has been omitted for simplicity.

Let

| | |
|---|---|
| $$Y(n) = Y_r(n) + jY_j(n)$$ | (A2) |
| $$X(n) = X_r(n) + jX_j(n)$$ | (A3) |
| $$g(l) = g_r(l) + jg_j(l)$$ | (A4) |

where subscript $r$ and $j$ denotes real and imaginary parts. The error $E$ can then be rewritten as

| | |
|---|---|
| $$E = \sum_{n=1}^{Nc} \left| \sum_{l=1}^{p+q+1} \left(g_r(l) + jg_j(l)\right)\left(X_r(n-l+q+1) + jX_j(n-l+q+1)\right) - \left(Y_r(n) + jY_j(n)\right) \right|^2$$ | (A5) |

Taking the derivates of the above with respect to $g_r(l)$ and $g_j(l)$ for each $l = 1, \ldots, p+q+1$ and setting them to zero leads to the following system of equations:

| | |
|---|---|
| $$\left(\boldsymbol{M_{rr}} + \boldsymbol{M_{jj}}\right)\boldsymbol{g_r} - \left(\boldsymbol{M_{rj}} - \boldsymbol{M_{jr}}\right)\boldsymbol{g_j} = \boldsymbol{R_{X_rY_r}} + \boldsymbol{R_{X_jY_j}}$$ | (A6) |
| $$\left(\boldsymbol{M_{jr}} - \boldsymbol{M_{rj}}\right)\boldsymbol{g_r} - \left(\boldsymbol{M_{rr}} + \boldsymbol{M_{jj}}\right)\boldsymbol{g_j} = \boldsymbol{R_{X_jY_r}} - \boldsymbol{R_{X_rY_j}}$$ | (A7) |

Where

$\boldsymbol{g}_r = [g_r(1), \ldots, g_r(q+1), \ldots, g_r(p+q+1)]$,

$\boldsymbol{g}_j = [g_j(1), \ldots, g_j(q+1), \ldots, g_j(p+q+1)]$,

$\boldsymbol{M_{rr}} = [a_{ij}]$ is a $(p+q+1)$x$(p+q+1)$ matrix with $a_{ij} = R_{X_rX_r}(i-j)$,

$\boldsymbol{M_{jj}} = [a_{ij}]$ is a $(p+q+1)$x$(p+q+1)$ matrix with $a_{ij} = R_{X_jX_j}(i-j)$,

$\boldsymbol{M_{rj}} = [a_{ij}]$ is a $(p+q+1)$x$(p+q+1)$ matrix with $a_{ij} = R_{X_rX_j}(i-j)$,

$\boldsymbol{M_{jr}} = [a_{ij}]$ is a $(p+q+1)$x$(p+q+1)$ matrix with $a_{ij} = R_{X_jX_r}(i-j)$,

$\boldsymbol{R_{X_rY_r}} = [a_i]$ is a $(p+q+1)$x1 vector with $a_i = R_{X_rY_r}(i-q-1)$,

$\boldsymbol{R_{X_jY_j}} = [a_i]$ is a $(p+q+1)$x1 vector with $a_i = R_{X_jY_j}(i-q-1)$,

$\boldsymbol{R_{X_rY_j}} = [a_i]$ is a $(p+q+1)$x1 vector with $a_i = R_{X_rY_j}(i-q-1)$,

$\boldsymbol{R_{X_jY_r}} = [a_i]$ is a $(p+q+1)$x1 vector with $a_i = R_{X_jY_r}(i-q-1)$,

$R_{X_rX_r}(i) = \sum_{n=i}^{N_c-1} X_r(n-i)X_r(n)$,

$R_{X_jX_j}(i) = \sum_{n=i}^{N_c-1} X_j(n-i)X_j(n)$,

$R_{X_rX_j}(i) = \sum_{n=i}^{N_c-1} X_r(n-i)X_j(n)$,

$R_{X_jX_r}(i) = \sum_{n=i}^{N_c-1} X_j(n-i)X_r(n)$,

$R_{X_rY_r}(i) = \sum_{n=i}^{N_c-1} X_r(n-i)Y_r(n)$,

$R_{X_jY_j}(i) = \sum_{n=i}^{N_c-1} X_j(n-i)Y_j(n)$,

$R_{X_rY_j}(i) = \sum_{n=i}^{N_c-1} X_r(n-i)Y_j(n)$, and

$R_{X_jY_r}(i) = \sum_{n=i}^{N_c-1} X_j(n-i)Y_r(n)$.

Solving the system of equations (A6) and (A7) leads to

| | |
|---|---|
| $\boldsymbol{g_r} = \left( (\boldsymbol{M_{rr}} + \boldsymbol{M_{jj}})^{-1} (\boldsymbol{M_{jr}} - \boldsymbol{M_{rj}}) - (\boldsymbol{M_{rj}} - \boldsymbol{M_{jr}})^{-1} (\boldsymbol{M_{rr}} + \boldsymbol{M_{jj}}) \right)^{-1} \left( (\boldsymbol{M_{rr}} + \boldsymbol{M_{jj}})^{-1} \left( \boldsymbol{R_{X_j Y_r}} - \boldsymbol{R_{X_r Y_j}} \right) - (\boldsymbol{M_{rj}} - \boldsymbol{M_{jr}})^{-1} \left( \boldsymbol{R_{X_r Y_r}} - \boldsymbol{R_{X_j Y_j}} \right) \right)$ | (A8) |
| $\boldsymbol{g_j} = \left( (\boldsymbol{M_{rr}} + \boldsymbol{M_{jj}})^{-1} (\boldsymbol{M_{rj}} - \boldsymbol{M_{jr}}) - (\boldsymbol{M_{jr}} - \boldsymbol{M_{rj}})^{-1} (\boldsymbol{M_{rr}} + \boldsymbol{M_{jj}}) \right)^{-1} \left( (\boldsymbol{M_{jr}} - \boldsymbol{M_{rj}})^{-1} \left( \boldsymbol{R_{X_j Y_r}} - \boldsymbol{R_{X_r Y_j}} \right) - (\boldsymbol{M_{rr}} + \boldsymbol{M_{jj}})^{-1} \left( \boldsymbol{R_{X_r Y_r}} - \boldsymbol{R_{X_j Y_j}} \right) \right)$ | (A9) |

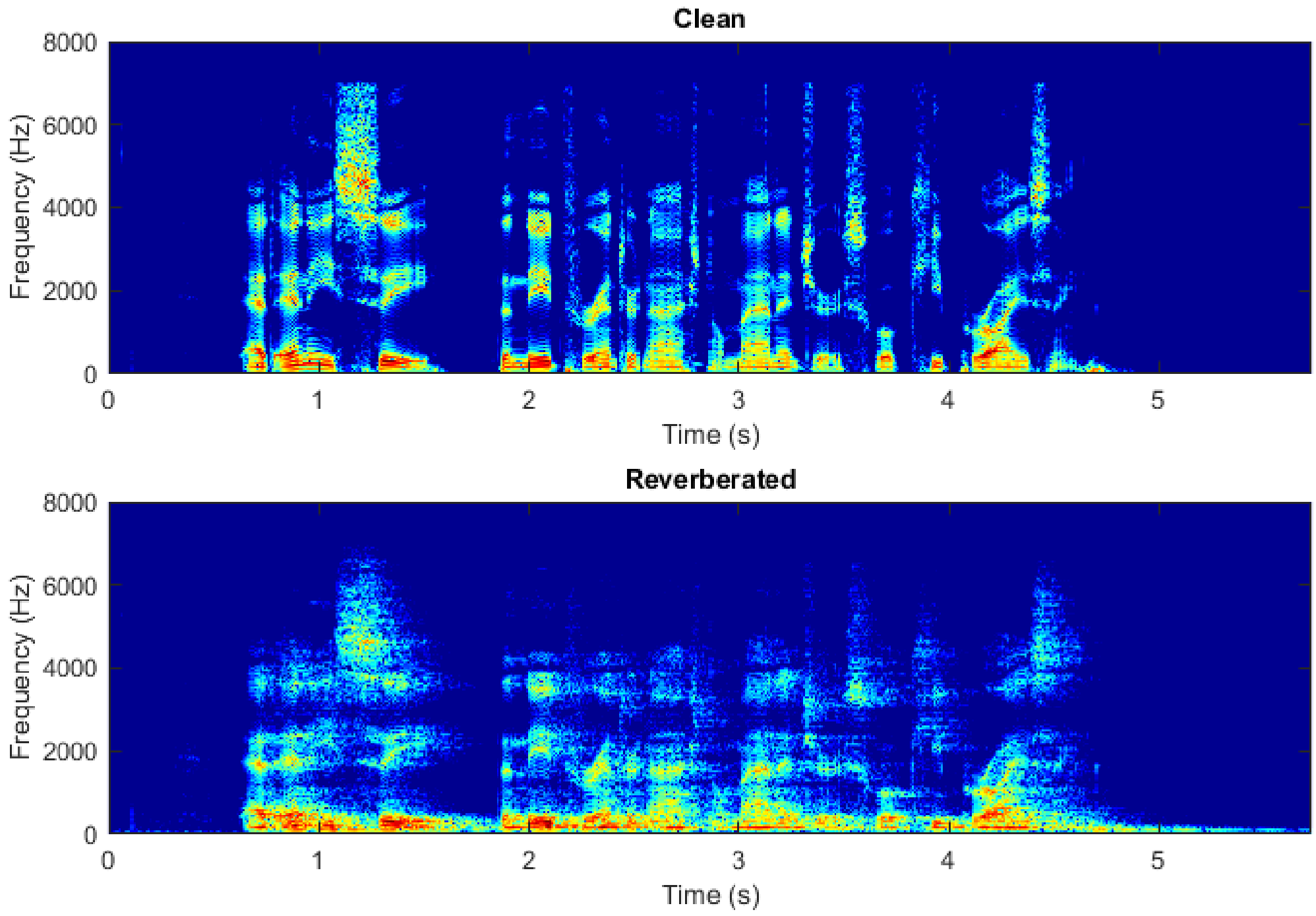

**Figure 1**: Spectrograms corresponding to an utterance before and after reverberation.

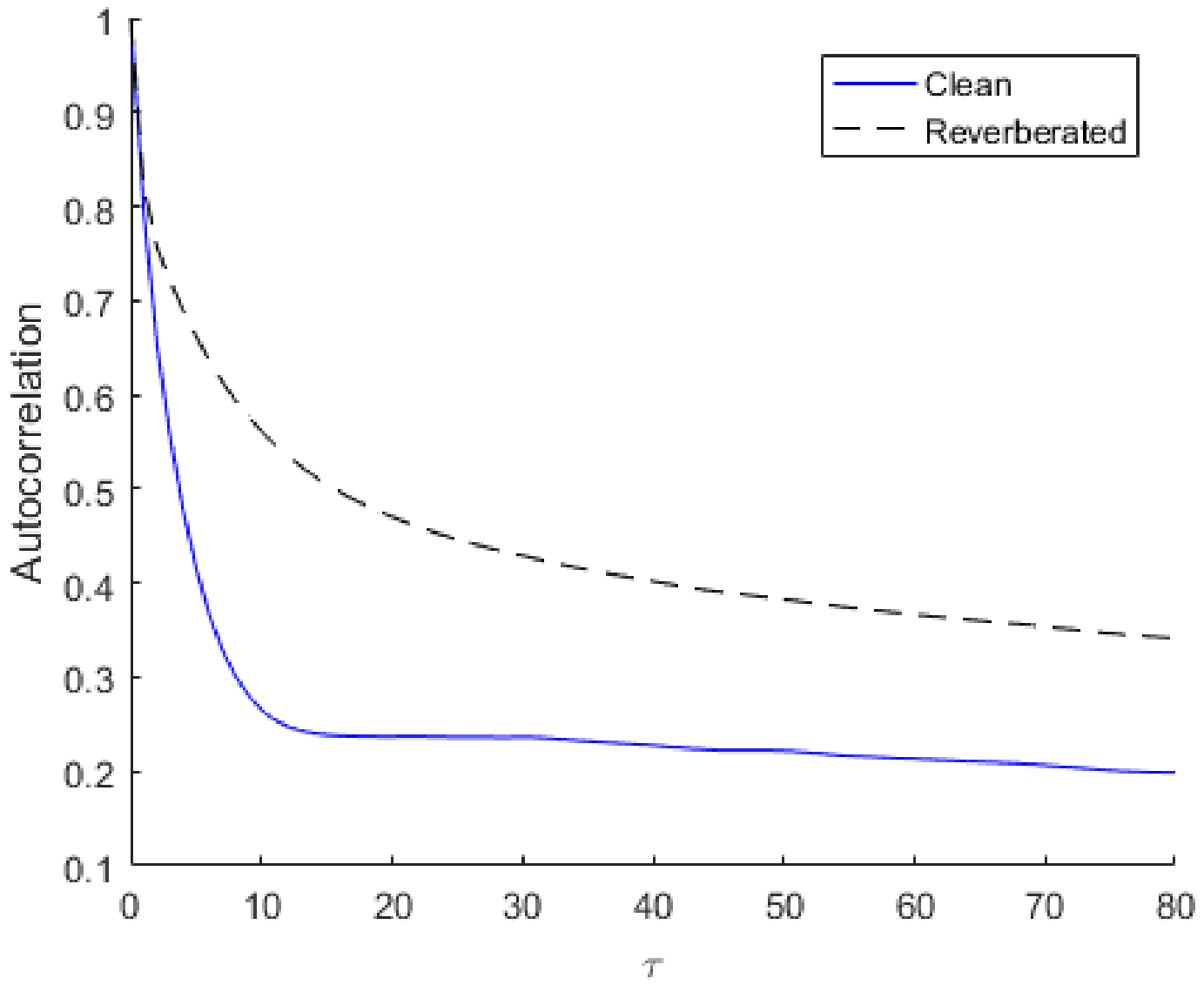

**Figure 2:** Average normalized autocorrelation of clean and reverberated testing utterances.

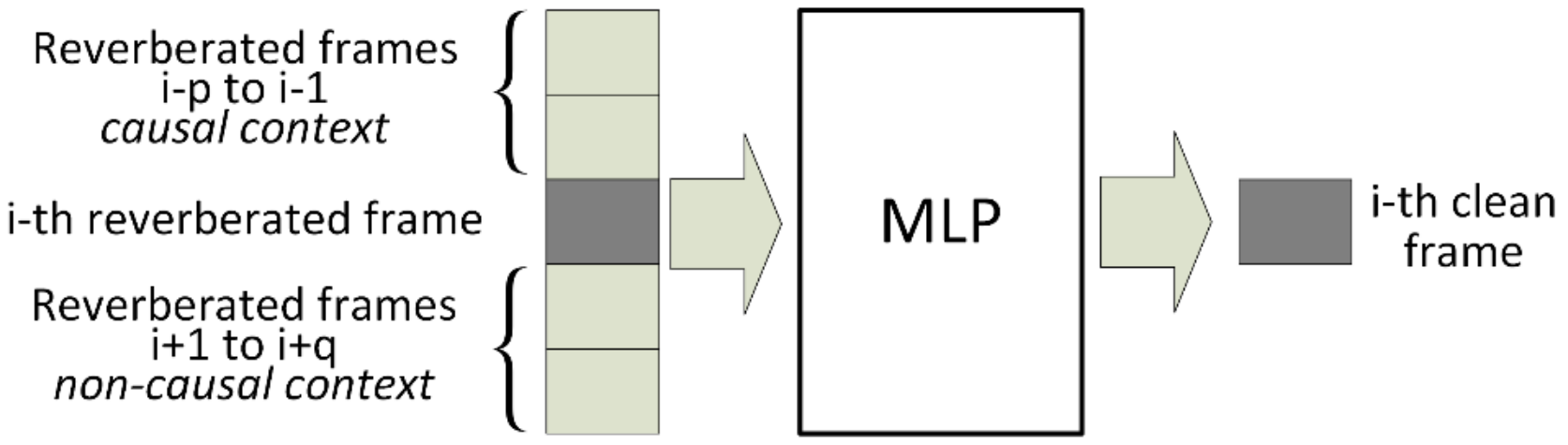

**Figure 3**: Non-causal MLP

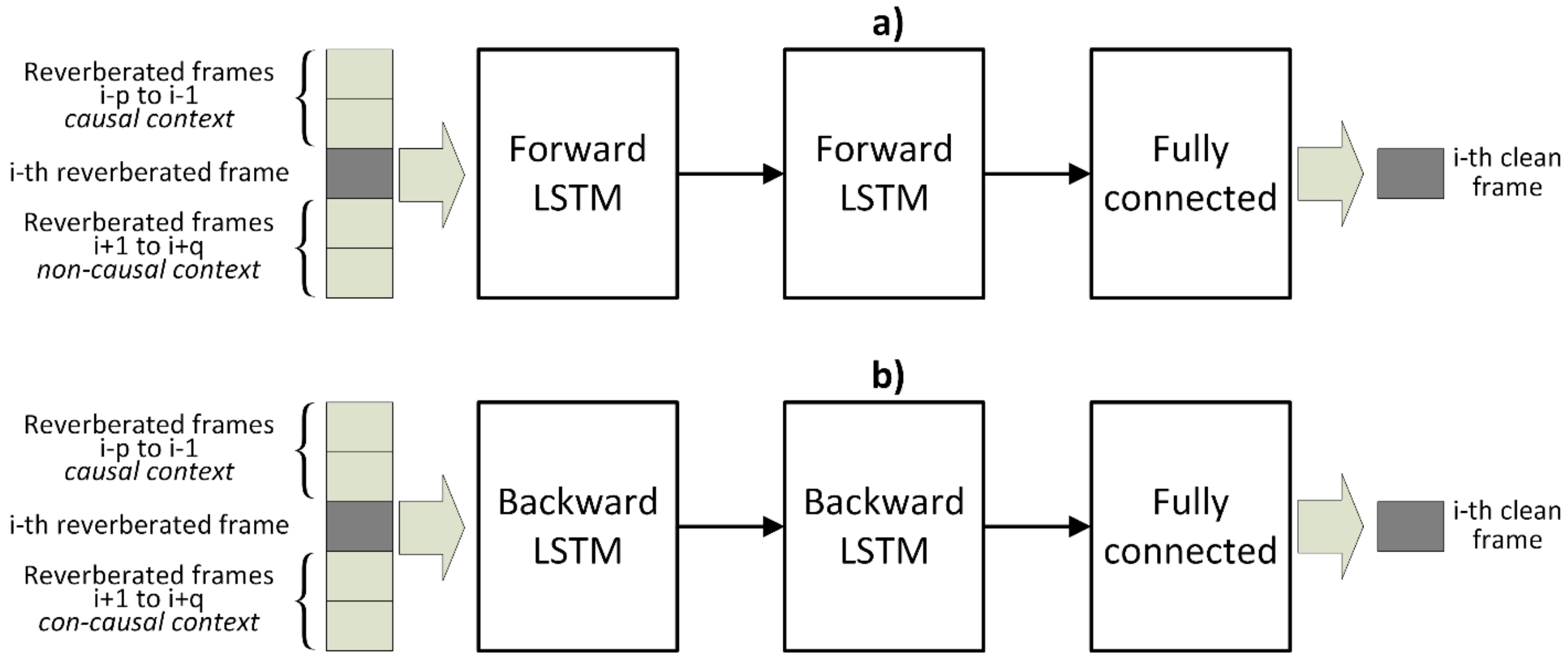

**Figure 4:** Non-causal unidirectional LSTM configurations: a) forward; b) backward.

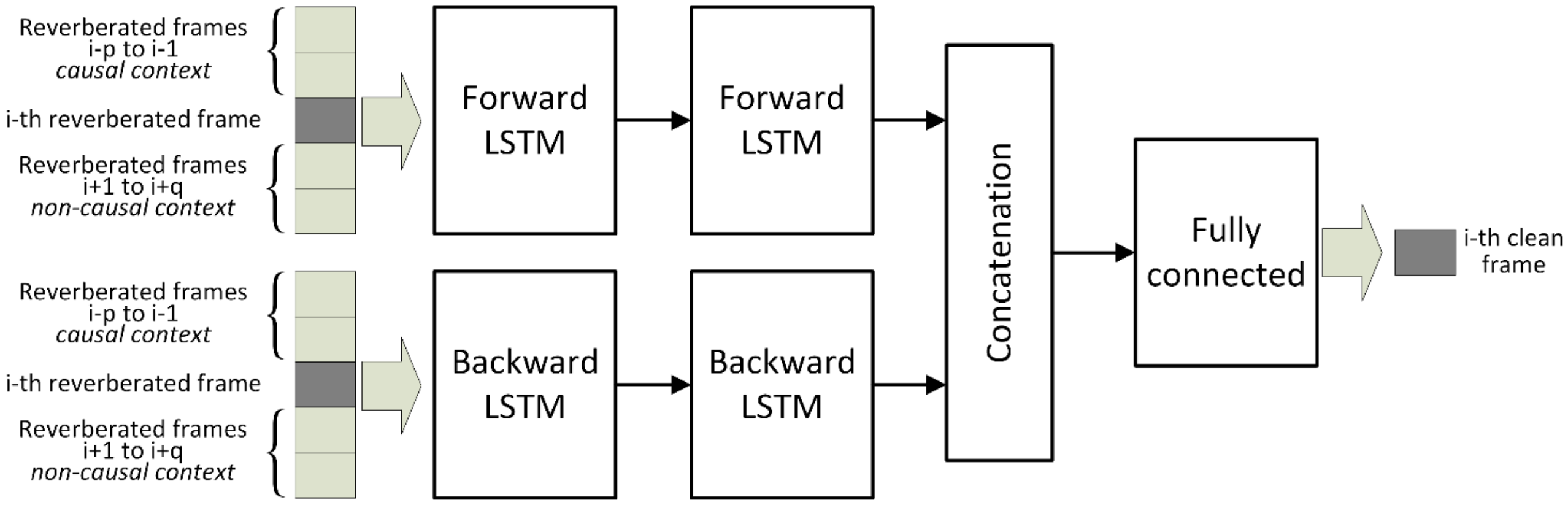

**Figure 5:** Non-causal bidirectional LSTM configuration.

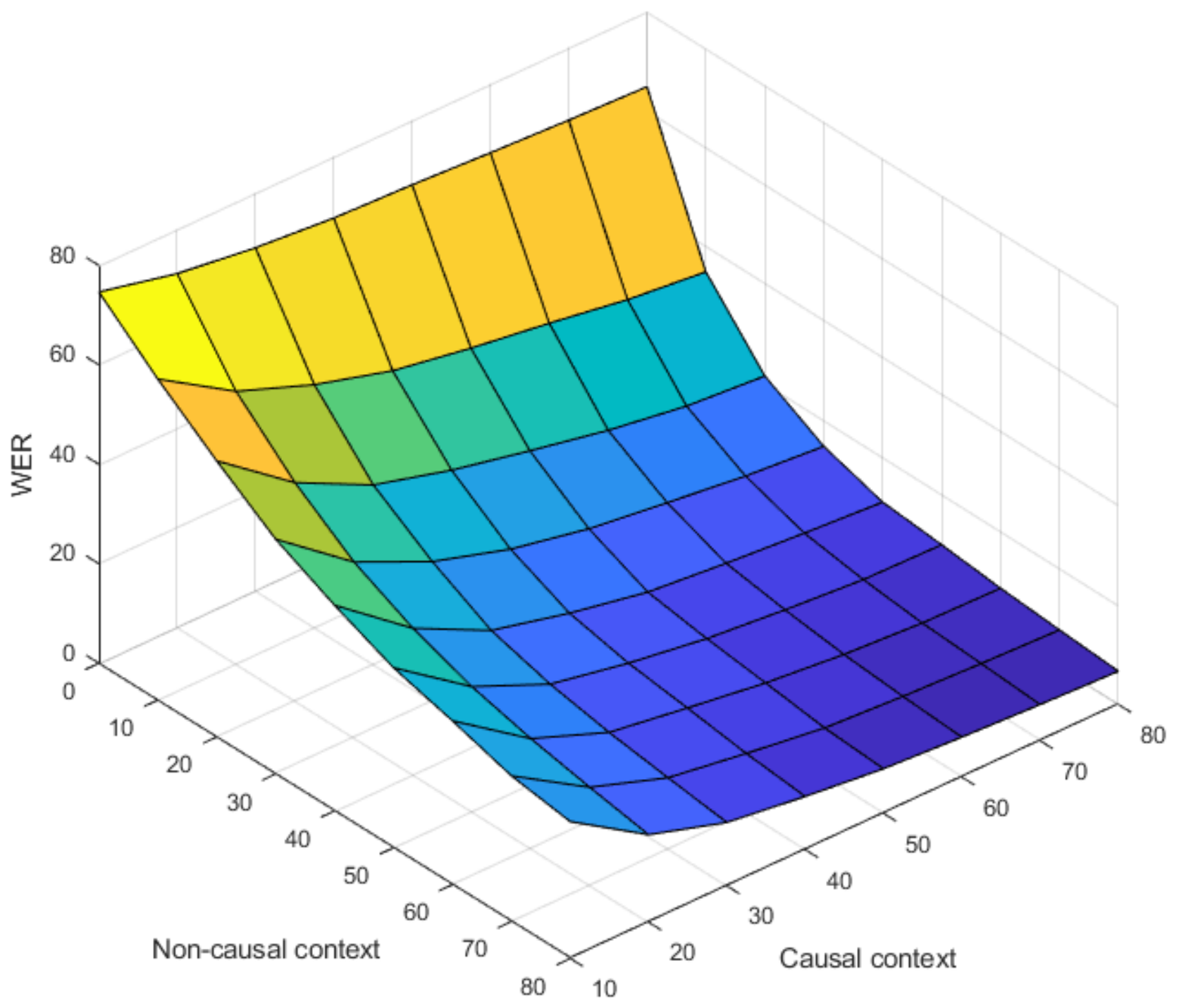

**Figure 6**: WER vs. FIR polynomial causal and non-causal contexts in Eq. (1).

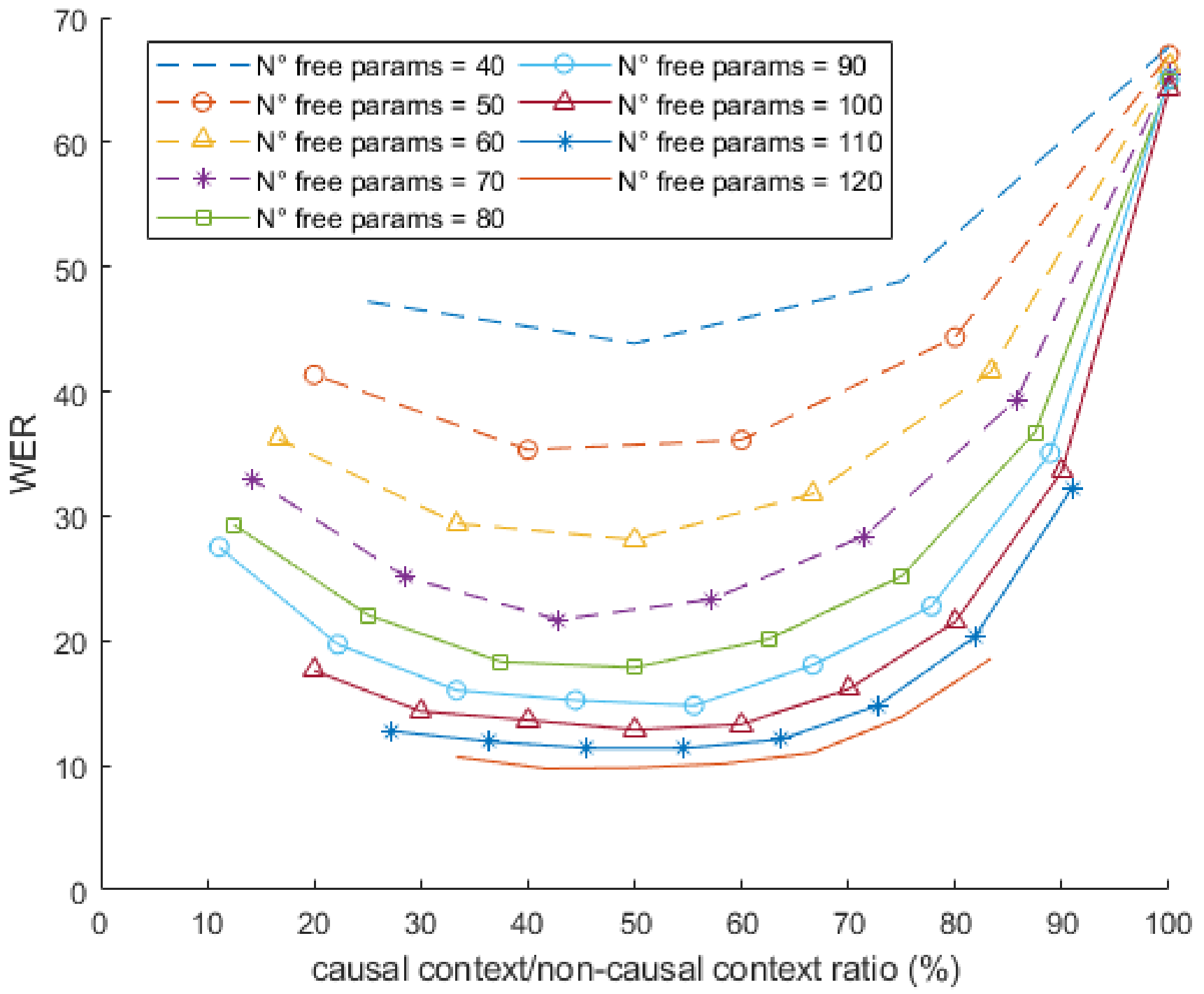

**Figure 7**: WER as a function of the causal context/non-causal context ratio (%). Each curve corresponds to a fixed number of free parameters in the FIR polynomial in Eq. (1).

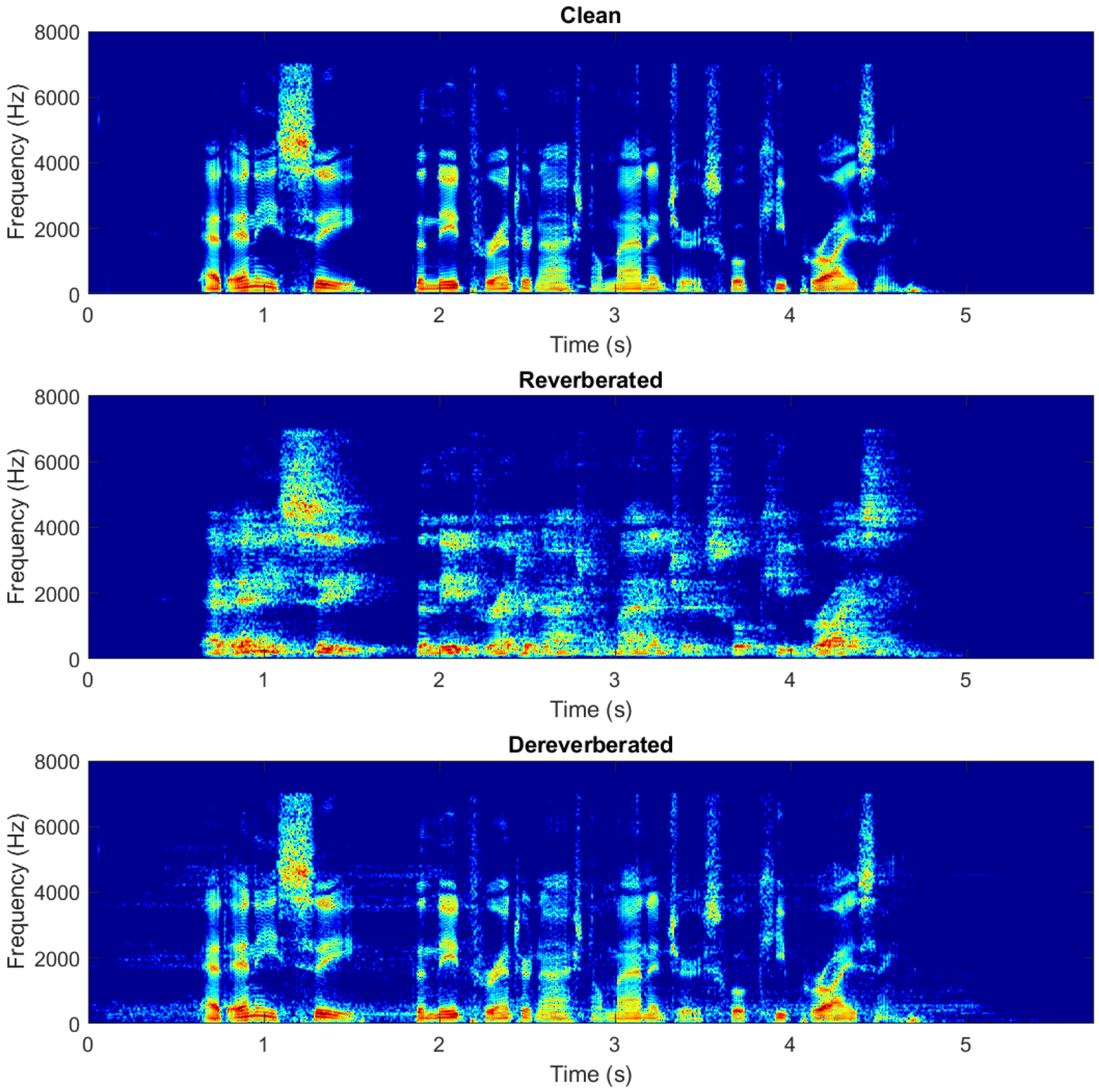

**Figure 8**: Spectrograms of a clean utterance and the corresponding reverberated and dereverberated signals. The dereverberated signal is estimated using the non-causal FIR polynomial.

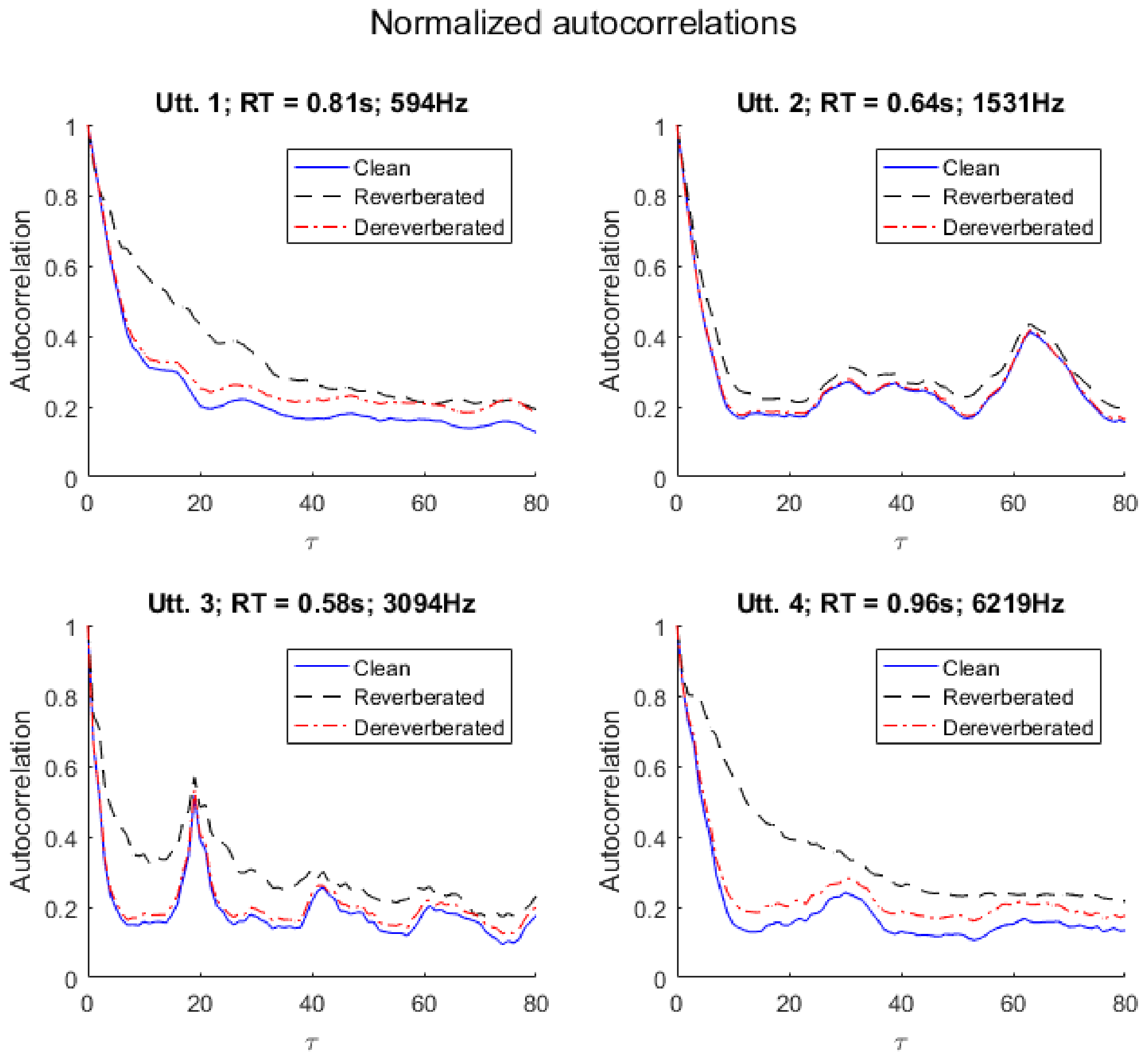

**Figure 9:** Normalized autocorrelations of clean, reverberated and dereverberated frequency bin trajectories.

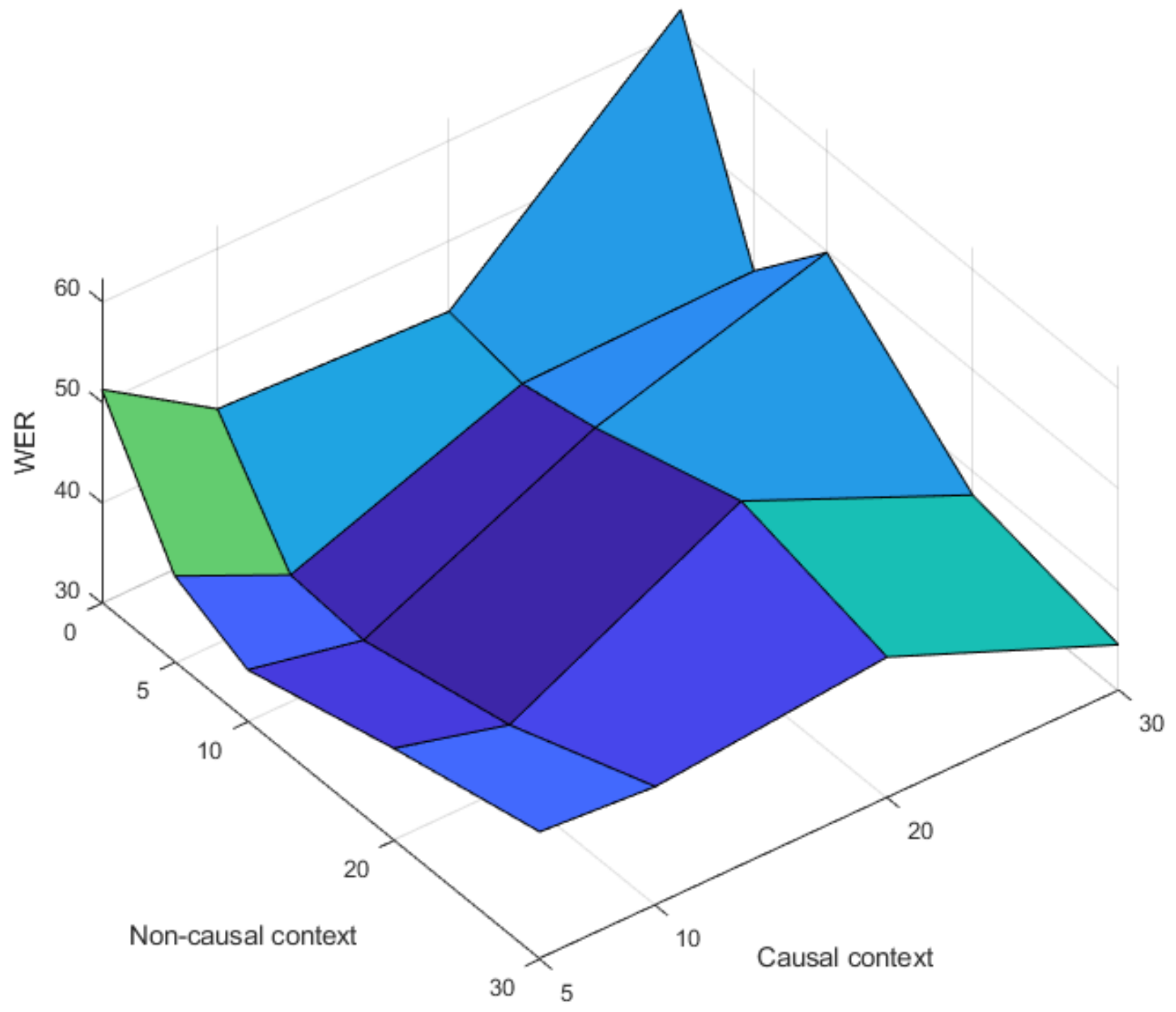

**Figure 10**: WER as a function of MLP causal and non-causal context.

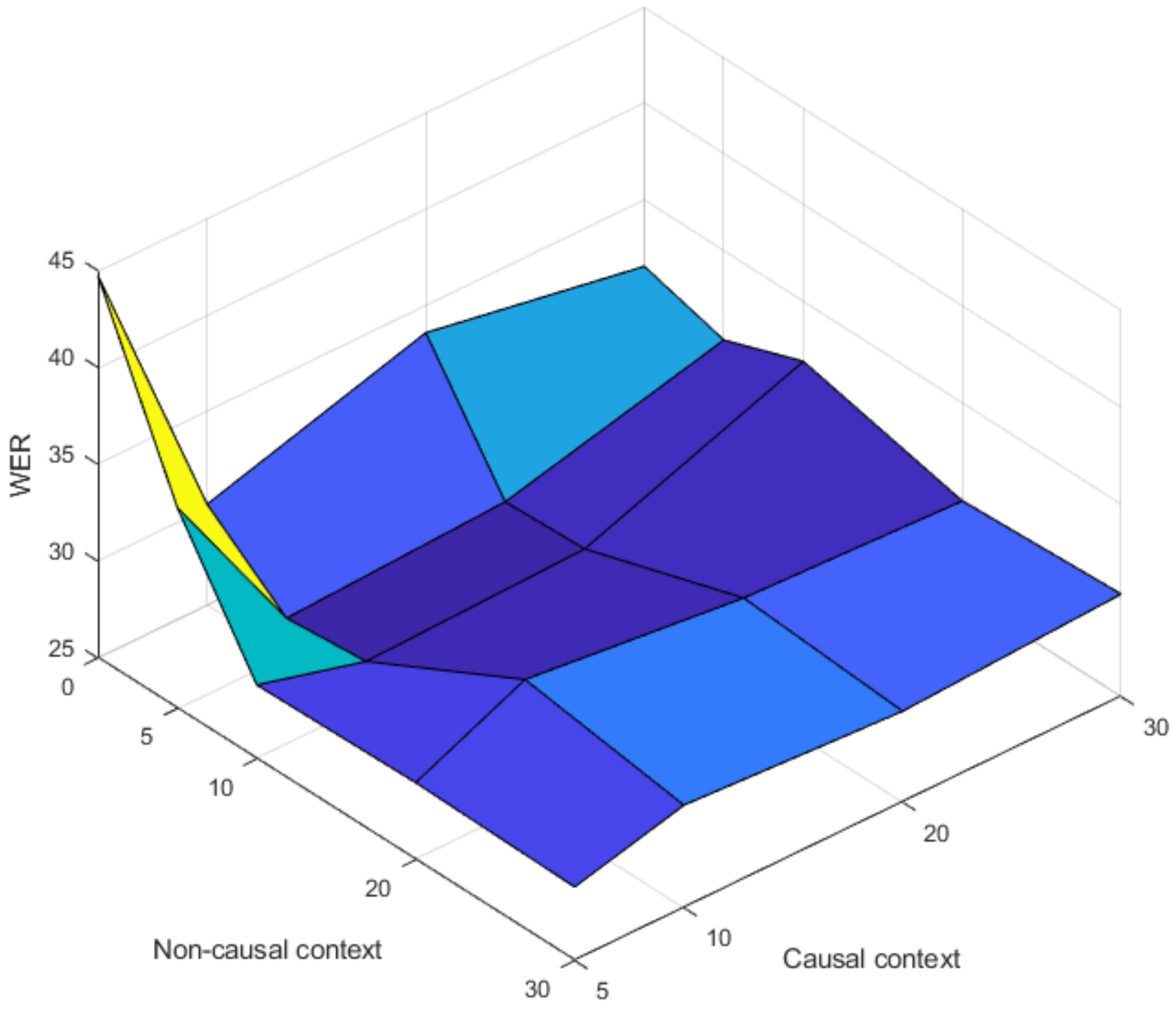

**Figure 11**: WER as a function of LSTM causal and non-causal context.

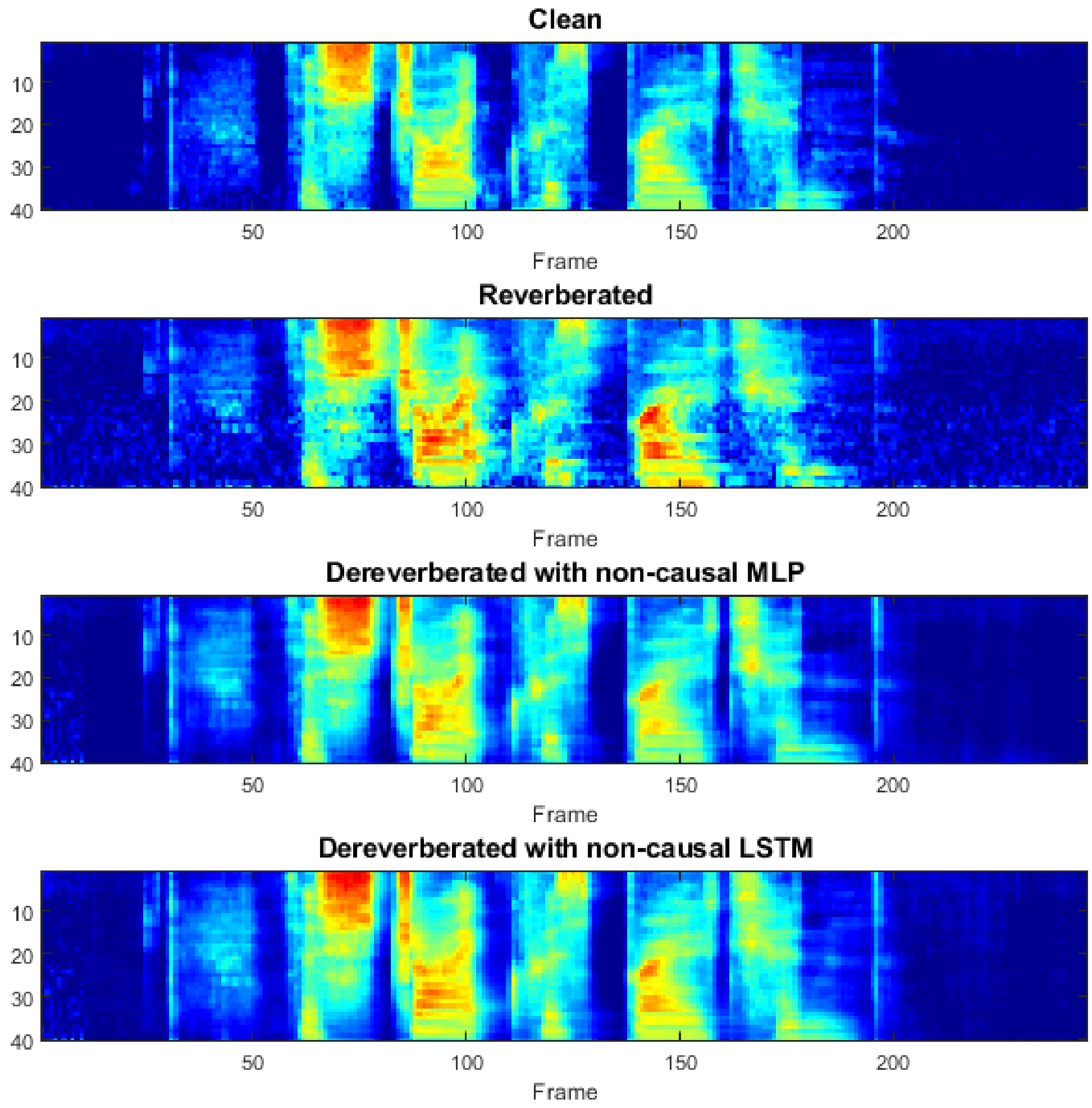

**Figure 12**: Log filter energy spectrograms of a clean speech segment, the corresponding reverberated signal, and the dereverberated signals obtained with the non-causal MLP and LSTM.

**Table 1.** WERs obtained from the AURORA-4 database using clean training. The number of causal and no-causal context were equal to 10 and 10 frames for MLP and 10 and 5 for LSTM, respectively.

| Clean | Reverb | WPE | MLP | LSTM |
|---|---|---|---|---|
| 2.19 | 76.63 | 58.55 | 32.77 | 26.94 |

**Table 2**. WERs obtained from the REVERB challenge database using clean training.

| Test | Clean | Reverb | WPE | MLP | LSTM |
|---|---|---|---|---|---|
| Real far | NA | 71.84 | 66.10 | 30.79 | 32.07 |
| Real near | NA | 70.78 | 65.83 | 27.60 | 28.04 |
| Simu far 1 | 3.50 | 12.54 | 10.02 | 7.96 | 7.25 |
| Simu far 2 | 4.10 | 57.29 | 50.45 | 16.90 | 16.95 |
| Simu far 3 | 4.02 | 69.51 | 64.89 | 19.33 | 18.52 |
| Simu near 1 | 3.50 | 6.62 | 5.93 | 6.57 | 5.84 |
| Simu near 2 | 4.10 | 20.01 | 15.87 | 9.57 | 8.99 |
| Simu near 3 | 4.02 | 28.24 | 23.13 | 10.02 | 9.78 |
| Avg. real tests | NA | 71.31 | 65.97 | 29.20 | 30.06 |
| Avg. simu tests | 3.50 | 41.66 | 37.93 | 17.78 | 17.65 |

**Table 3.** WERs obtained from the REVERB challenge database using MCT

| Test | Reverb | WPE | MLP | LSTM |
|---|---|---|---|---|
| Real far | 17.69 | 16.68 | 20.56 | 22.38 |
| Real near | 15.94 | 14.12 | 19.71 | 19.48 |
| Simu far 1 | 4.33 | 4.18 | 7.35 | 7.13 |
| Simu far 2 | 7.58 | 7.1 | 10.99 | 12.45 |
| Simu far 3 | 8.37 | 7.7 | 12.24 | 13.46 |
| Simu near 1 | 3.61 | 3.57 | 6.26 | 5.99 |
| Simu near 2 | 4.85 | 4.9 | 6.94 | 7.44 |
| Simu near 3 | 5.5 | 5.21 | 7.26 | 7.63 |
| Avg. real tests | 16.82 | 15.40 | 20.14 | 20.93 |
| Avg. simu tests | 10.14 | 9.15 | 13.53 | 13.31 |

**Table 4.** WERs obtained from the REVERB challenge database using expanded MCT. The non-causal MLP was employed for dereverberation.

| Test | Reverb | WPE | MLP |
|---|---|---|---|
| Real far | 19.38 | 18.43 | 18.57 |
| Real near | 17.60 | 15.46 | 16.96 |
| Simu far 1 | 3.94 | 3.61 | 4.44 |
| Simu far 2 | 8.61 | 7.89 | 8.24 |
| Simu far 3 | 9.03 | 8.18 | 8.25 |
| Simu near 1 | 3.30 | 3.27 | 3.71 |
| Simu near 2 | 5.16 | 5.04 | 5.04 |
| Simu near 3 | 5.30 | 5.10 | 5.16 |
| Avg. real tests | 18.49 | 16.95 | 17.77 |
| Avg. simu tests | 10.77 | 9.54 | 10.70 |

**Table 5.** WERs obtained from the REVERB challenge database using expanded MCT. The non-causal LSTM was employed for dereverberation.

| Test | Reverb | WPE | LSTM |
|---|---|---|---|
| Real far | 18.74 | 17.29 | 19.14 |
| Real near | 16.19 | 14.76 | 17.37 |
| Simu far 1 | 4.15 | 3.78 | 4.11 |
| Simu far 2 | 8.16 | 7.52 | 7.92 |
| Simu far 3 | 8.93 | 8.38 | 8.47 |
| Simu near 1 | 3.17 | 3.11 | 3.59 |
| Simu near 2 | 4.45 | 4.51 | 4.75 |
| Simu near 3 | 5.39 | 5.23 | 4.74 |
| Avg. real tests | 17.47 | 16.03 | 18.26 |
| Avg. simu tests | 10.17 | 9.27 | 10.74 |

**Table 6**. WERs obtained from semi-enhanced test utterances combined with expanded MCT. The optimal $\boldsymbol{\lambda}$ is subset dependent. The non-causal MLP was employed for dereverberation.

| | **Expanded MCT** | | | | **MCT** | |
|---|---|---|---|---|---|---|
| **Test** | **Conf 1** | **Conf 2** | **Conf 3** | **Conf 4** | **Reverb** | **WPE** |
| Real far | 16.91 | 17.25 | 17.29 | 17.49 | 17.69 | 16.68 |
| Real near | 14.79 | 14.72 | 15.46 | 16.45 | 15.94 | 14.12 |
| Simu far 1 | 3.57 | 3.54 | 3.76 | 3.86 | 4.33 | 4.18 |
| Simu far 2 | 7.73 | 7.24 | 7.34 | 7.95 | 7.58 | 7.1 |
| Simu far 3 | 7.86 | 7.58 | 7.29 | 8.44 | 8.37 | 7.7 |
| Simu near 1 | 3.23 | 3.23 | 3.25 | 3.25 | 3.61 | 3.57 |
| Simu near 2 | 5.01 | 5.04 | 4.87 | 4.95 | 4.85 | 4.90 |
| Simu near 3 | 4.94 | 5.21 | 5.03 | 5.08 | 5.5 | 5.21 |
| Avg. real tests | 15.85 | 15.99 | 16.38 | 16.97 | 16.82 | 15.40 |
| Avg. simu tests | 9.18 | 9.13 | 9.61 | 10.16 | 10.14 | 9.15 |

**Table 7**. WERs obtained from semi-enhanced test utterances combined with expanded MCT. The optimal $\boldsymbol{\lambda}$ is averaged across the subsets. The non-causal MLP was employed for dereverberation.

| | **Expanded MCT** | | | | **MCT** | |
|---|---|---|---|---|---|---|
| **Test** | **Conf 1** | **Conf 2** | **Conf 3** | **Conf 4** | **Reverb** | **WPE** |
| Real far | 16.91 | 16.91 | 17.32 | 17.59 | 17.69 | 16.68 |
| Real near | 14.79 | 14.21 | 15.39 | 15.91 | 15.94 | 14.12 |
| Simu far 1 | 3.57 | 3.69 | 3.78 | 3.86 | 4.33 | 4.18 |
| Simu far 2 | 7.37 | 7.34 | 7.42 | 7.57 | 7.58 | 7.1 |
| Simu far 3 | 7.29 | 7.41 | 7.75 | 7.82 | 8.37 | 7.7 |
| Simu near 1 | 3.28 | 3.27 | 3.25 | 3.22 | 3.61 | 3.57 |
| Simu near 2 | 5.01 | 4.96 | 4.87 | 4.9 | 4.85 | 4.9 |
| Simu near 3 | 5.03 | 4.93 | 5.04 | 4.98 | 5.5 | 5.21 |
| Avg. real tests | 15.85 | 15.56 | 16.36 | 16.75 | 16.82 | 15.40 |
| Avg. simu tests | 9.18 | 8.95 | 9.59 | 9.89 | 10.14 | 9.15 |

**Table 8**. WERs obtained from semi-enhanced test utterances combined with expanded MCT. The optimal $\boldsymbol{\lambda}$ is subset dependent. The non-causal LSTM was employed for dereverberation.

| | **Expanded MCT** | | | | **MCT** | |
|---|---|---|---|---|---|---|
| **Test** | **Conf 1** | **Conf 2** | **Conf 3** | **Conf 4** | **Reverb** | **WPE** |
| Real far | 16.95 | 16.64 | 18.23 | 17.56 | 17.69 | 16.68 |
| Real near | 14.08 | 14.24 | 14.44 | 15.01 | 15.94 | 14.12 |
| Simu far 1 | 3.84 | 3.91 | 3.79 | 3.98 | 4.33 | 4.18 |
| Simu far 2 | 7.26 | 7.00 | 6.97 | 7.12 | 7.58 | 7.1 |
| Simu far 3 | 7.92 | 7.17 | 7.58 | 7.94 | 8.37 | 7.7 |
| Simu near 1 | 3.11 | 3.28 | 3.39 | 3.32 | 3.61 | 3.57 |
| Simu near 2 | 4.43 | 4.43 | 4.3 | 4.4 | 4.85 | 4.9 |
| Simu near 3 | 4.7 | 4.84 | 4.99 | 4.87 | 5.5 | 5.21 |
| Avg. real tests | 15.52 | 15.44 | 16.34 | 16.29 | 16.82 | 15.40 |
| Avg. simu tests | 8.96 | 9.08 | 9.12 | 9.50 | 10.14 | 9.15 |

**Table 9**. WERs obtained from semi-enhanced test utterances combined with expanded MCT. The optimal $\boldsymbol{\lambda}$ is averaged across the subsets. The non-causal LSTM was employed for dereverberation.

| | **Expanded MCT** | | | | **MCT** | |
|---|---|---|---|---|---|---|
| **Test** | **Conf 1** | **Conf 2** | **Conf 3** | **Conf 4** | **Reverb** | **WPE** |
| Real far | 16.68 | 16.44 | 16.91 | 17.18 | 17.69 | 16.68 |
| Real near | 14.34 | 13.77 | 14.79 | 15.01 | 15.94 | 14.12 |
| Simu far 1 | 3.86 | 3.76 | 3.89 | 4.03 | 4.33 | 4.18 |
| Simu far 2 | 7.15 | 6.99 | 7.02 | 7.12 | 7.58 | 7.1 |
| Simu far 3 | 7.57 | 7.12 | 7.58 | 7.74 | 8.37 | 7.7 |
| Simu near 1 | 3.20 | 3.23 | 3.25 | 3.25 | 3.61 | 3.57 |
| Simu near 2 | 4.42 | 4.37 | 4.19 | 4.5 | 4.85 | 4.9 |
| Simu near 3 | 5.11 | 4.96 | 4.99 | 4.87 | 5.5 | 5.21 |
| Avg. real tests | 15.51 | 15.11 | 15.85 | 16.10 | 16.82 | 15.40 |
| Avg. simu tests | 9.10 | 8.77 | 9.34 | 9.52 | 10.14 | 9.15 |